\documentclass{elsart_jihj}
 \usepackage{graphicx}
 \usepackage{t1enc}

\newcommand{\mssb}{M\"{o}ss\-bauer}
\newcommand{\lande}{Land\'{e}}
\newcommand{\fit}{\textsf{Fit;o)}}
\newcommand{\Fe}{$^{\rm 57}$Fe}

\newcommand{\beq}{\begin{equation}}
\newcommand{\eeq}{\end{equation}}
\newcommand{\vb}[1]{\texttt{#1}} %verbatim bold
\newcommand{\pb}[1]{\texttt{#1}} %programmiong bold
\newcommand{\fb}[1]{\texttt{#1}} % physics bold
\newcommand{\ee}[1]{\cdot 10^{#1}}
\newcommand{\eref}[1]{(\ref{#1})}
\newcommand{\fref}[1]{figure \ref{#1}}
\newcommand{\sref}[1]{section \ref{#1}}

\newcommand{\tref}[1]{table \ref{#1}}

\newcommand{\Lorentzian}{Lorentzian}
\newcommand{\Gaussian}{Gaussian}
\newcommand{\SplitLorentzian}{Split-Lorentzian}
\newcommand{\PseudoVoigt}{Pseudo-Voigt}
\newcommand{\PseudoLorentz}{Pseudo-Lorentz}
\newcommand{\PearsonVII}{Pearson-VII}

\newcommand{\ProgramVersion}{1.0.0.63}

\newcounter{bla}

\begin{document}
\begin{frontmatter}

\title{\fit\ - A \mssb\ spectrum fitting program}

\author[a,b]{Jari \'{i} Hj\o{}llum},
\ead{jari.hjoellum@risoe.dk}
\author[a]{Morten Bo Madsen},
\ead{mbmadsen@fys.ku.dk}

\address[a]{Earth and Planetary Physics, Juliane Maries vej 30, Niels Bohr Institute, University of Copenhagen,
Denmark.}

\address[b]{Materials Research Department, Building 775, Ris\o{} National Laboratory, Technical University of Denmark, Fredriksborgvej 399, DK-4000 Roskilde, Denmark.}

\begin{abstract}
\fit\ is a \mssb\ fitting and analysis program written in Borland
Delphi. It has a complete graphical user interface that allows all
actions to be carried out via mouse clicks or key shortcut
operations in a WYSIWYG fashion.  The program does not perform
complete transmission integrals, and will therefore not be suited
for a complete analysis of all types of \mssb\ spectra and e.g. low
temperature spectra of ferrous silicates. Instead, the program is
intended for application on complex spectra resulting from typical
mineral samples, in which many phases and different crystallite
sizes are often present at the same time. The program provides the
opportunity to fit the spectra with \Gaussian{}, \Lorentzian{},
\SplitLorentzian{}, \PseudoVoigt{}, \PseudoLorentz{} and
\PearsonVII{} line profiles for individual components of the
spectra. This feature is particularly useful when the sample
contains components, that are affected by effects of either
relaxation or interaction among particles. Fitted spectra may be
printed, fits saved, data files exported for graph creation in other
programs, and analysis tables and reports may be exported as plain
text or \LaTeX{} files. With \fit{} even an inexperienced user will
soon be able to analyze and fit relatively complex \mssb\ spectra of
mineralogical samples quickly without programming knowledge.

\begin{flushleft}
  %Insert your suggested PACS number here

\end{flushleft}

\begin{keyword}

\mssb\ spectra, \mssb\ fitting, mineralogical analysis, \Fe{},
complex mixture

% Please give some freely chosen keywords that we can use in a
  % cumulative keyword index.
\end{keyword}

\end{abstract}

\end{frontmatter}

\section{Introduction}\label{sec:Introduction}
During work with analysis of \mssb\ spectra we found that
 existing analysis software packages however
competent (i.e. Mfit \cite{mfit}, Recoil \cite{recoil}, MacFit
\cite{macfit}, MossWinn\cite{mosswinn}) could not fulfill our needs
of being simple and yet very flexible. For fitting of spectra with
many components, as is typical for samples with complex mineral
assemblies, advanced programs are in many cases too detailed to be
really useful.

We therefore developed this program package which attempts to
mitigate these issues, and thus provides a powerful tool for rapid
fitting of complex \mssb\ spectra.

The program has been downloaded around 100 times in several
versions, and an estimate of regular users is around 20. The program
package has been used in the preparation of several articles or
talks
\cite{CarpenterKuhn,Duprat,HjoellumMagnetite,GoetzMars7Conf2007,MadsenJGR2007}.

\section{Program summary}\label{sec:ProgramSummary}
\fit\ is a program for fitting and analyzing transmission and
scattering geometry \Fe\ \mssb\ spectra of metals, alloys and
mixtures of ferric oxides, oxyhydrates and silicates. The program
can be expanded to handle other \mssb\ isotopes as well. It has a
complete graphical user interface, which allows all actions to be
performed via mouse clicks or key shortcuts. The program accepts
currently both particular plain text files (.exp-files, see
\cite{FitReport,FitWebpage} for a definition) and comma-separated
files e.g. counts vs. velocity (velocity;counts) as data input. It
is the intention to release the source code under an open license.

\mssb\ spectra can be fitted using singlets, doublets and sextets
(all with a selection of line profiles of which \Lorentzian{} is
default) and the fit model can be saved for later reloading for
example for further refinement of the fit. Fitted and unfitted
spectra can be saved as reports and saved as both plain text and in
customizable \LaTeX\-format, and spectra can be printed with a fit
report.

The program contains a precompiled list of \mssb{} data of many
common iron compounds with an option for the user to edit or add new
compounds.

Calibration data for \mssb\ experimental setups can be extracted
from calibration spectra, and saved in separate calibration
.cal-files (see \cite{FitReport,FitWebpage} for a description).

The settings window contains a wide range of customizable settings
through which the user may customize the appearance and behavior of
the program. Backwards compatibility is ensured, as file formats are
kept unchanged and user-changed files are not overwritten.

The web update option provides an easy and automated way to keep the software updated with the latest program version.

The program does not perform complete transmission integrals, and
will therefore not be suited for a complete analysis of all types of
\mssb{} spectra. Also the fitting does not use the complete
spin-Hamiltonian (but approximations) so for example magnetic
(low-temperature) spectra of silicates cannot be analyzed by the
programme.

Some of \fit{}'s main features are:
\begin{itemize}
    \item Microsoft Windows 2000/XP compatible.
    \item Easy to install, no external dependencies, and safe uninstallation/removal.
    \item Complete point-and-click graphical user interface.
    \item Easy saving and loading of fit models.
    \item Most program parameters as customizable.
    \item Working with several spectra (MDI\footnote{Multi Document Interface}) at a time is possible.
    \item Created with object-oriented programming.

%    \item Easy and safe installation and uninstallation/removal.
\end{itemize}

\section{Description}\label{sec:Description}
\subsection{Prerequisites}\label{sec:DesPrerequisites}
As noted in the introduction, execution of \fit\ requires an updated
Microsoft Windows 2000, Microsoft Windows XP operating system. A
screen resolution of a least 1024x768 with a 16 bit color depth is
recommended.

\subsection{Installation and execution}\label{sec:DesInstallExecution}
The program is available at \pb{http://hjollum.com/jari/zzbug/fit/}.
After installation and execution the main options are:

\begin{itemize}
  \item Open the selected type spectrum by selecting  an appropriate
  icon in \pb{File} menu.
  \item To open a file in text mode, use the \vb{Open text file} option and to view the
    application log, use the \vb{Open log} option.
\end{itemize}
A more detailed description of the program interface can be found in
the manual \cite{FitReport} or the web site \cite{FitWebpage}.

\subsection{The layout of the graphical user
interface}\label{sec:DesLayoutGraph}

\begin{figure}[!htb]
\centering
\includegraphics[width=0.6\textwidth]{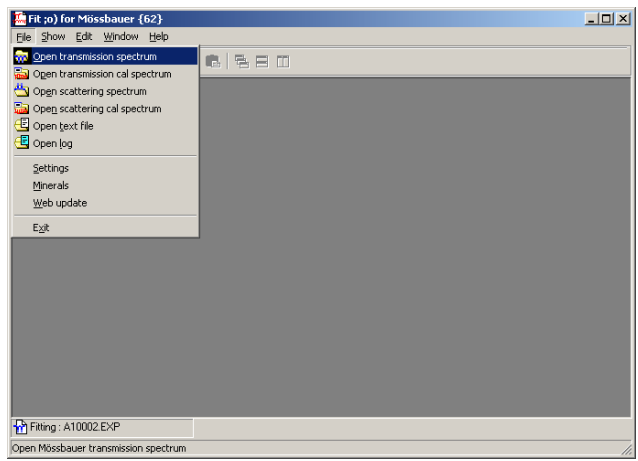}
\caption{\textit{Opening a spectrum file.}} \label{fig:OpenFitForm}
\end{figure}

After the first execution of the program the user is presented to
the graphical user interface shown in \fref{fig:OpenFitForm}. The
main window of \fit\ is divided into five areas (listed from top
down), the menu bar, the tool bar, the work area, the window panel
and the status panel.

\subsection{Fitting}\label{sec:FittingFormSection}

\begin{figure}[!htb]
\centering
\includegraphics[width=0.6\textwidth]{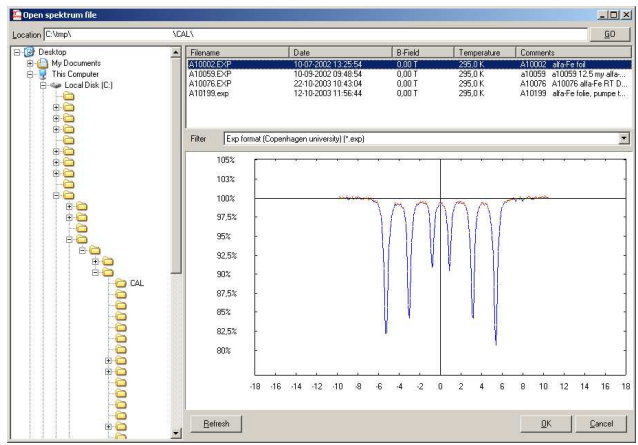}
\caption{\textit{Choosing the spectrum to open.}}
\label{fig:OpenSpectrumForm}
\end{figure}

\begin{figure}[!htb]
\centering
\includegraphics[width=0.6\textwidth]{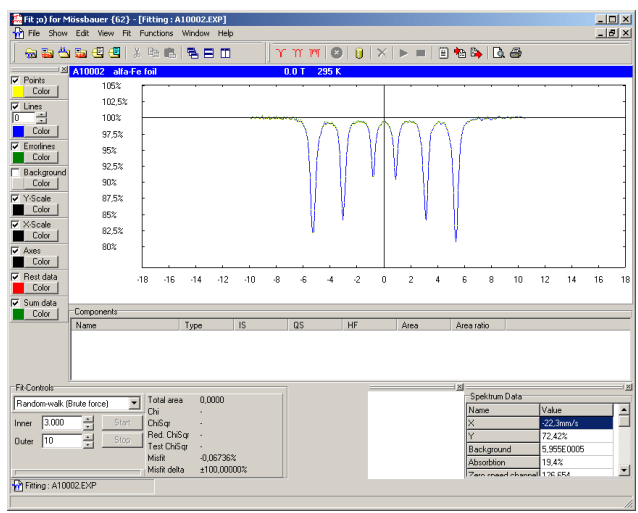}
\caption{\textit{The fit form with a loaded spectrum.}}
\label{fig:FitFormWindow}
\end{figure}

When choosing to open a spectrum for fitting, the open spectrum form
(\fref{fig:OpenSpectrumForm}) appears, in which the spectrum to be
loaded is chosen. After choosing a spectrum the fit form
(\fref{fig:FitFormWindow}) is opened and the spectrum is loaded.

As fitting components, it is possible to insert either simple
(fundamental) components (singlets, doublets, sextets), or to insert
predefined mineral components.

Simple components are inserted by clicking one of the spectrum icons
in the toolbar, and then marking, position, intensity and width of
the components on the graph. The parameter that the user is expected
to mark is indicated at the cursor.

Predefined (but editable) mineral components are inserted by
clicking the cylinder icon in the toolbar, and choosing the mineral.
After choosing, one has to mark the intensity of the component on
the graph using the mouse cursor.

\begin{figure}[!htb]
\centering
\includegraphics[width=0.6\textwidth]{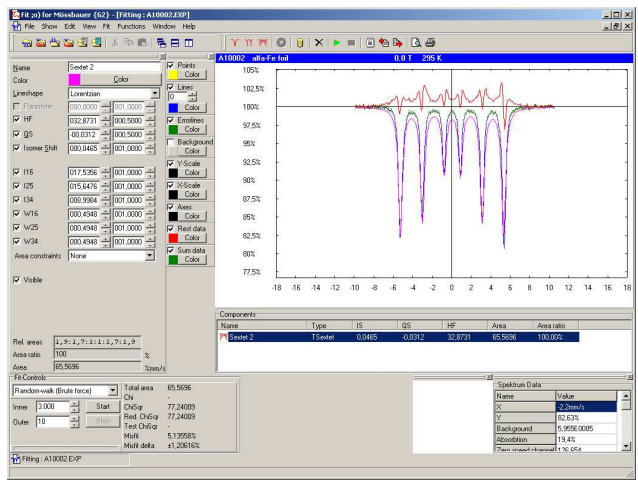}
\caption{\textit{The fit form with an inserted component.}}
\label{fig:FitFormActive}
\end{figure}

The fitting algorithm is chosen in the \vb{Fit Controls} combobox in
the lower left corner of the form. The fitting process is started
and aborted by using the adjacent start and stop.

\subsection{Calibration}\label{sec:CalibrationFormSection}

\begin{figure}[!htb]
\centering
\includegraphics[width=0.4\textwidth]{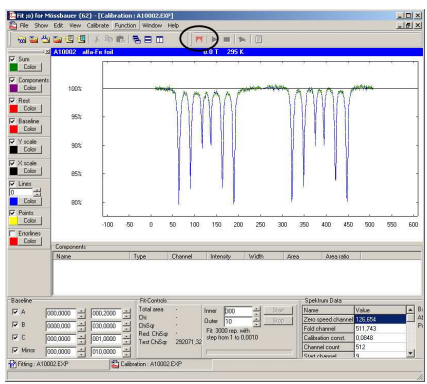}
\caption{\textit{The calibration form before inserting the calibration fit components. The insert button is marked.}}
\label{fig:CalBeforeInsComp}
\end{figure}

Fitting components are inserted by clicking the button in the
toolbar with the sextet icon (see \fref{fig:CalBeforeInsComp}).
\begin{figure}[!htb]
\centering
\includegraphics[width=0.4\textwidth]{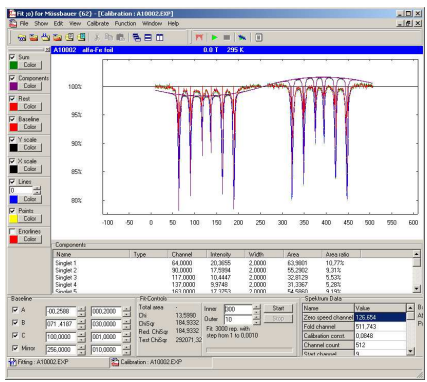}
\caption{\textit{The calibration form after insertion of the
calibration fit components.}} \label{fig:CalInsComp}
\end{figure}
When the components have been inserted, either a single fitting run
or a series of repetitive fitting runs can be started, through the
start icons. The result can be saved and exported by clicking the
'notepad' icon.

%the fitting process can be started either . By clicking the either
%the play icon (or \vb{Start} button), which runs a single fitting
%run, or by clicking the "play with lightning" icon (second from
%right), which starts a repetitive fitting process.

\subsection{Minerals}
\begin{figure}[!htb]
\centering
\includegraphics[width=0.6\textwidth]{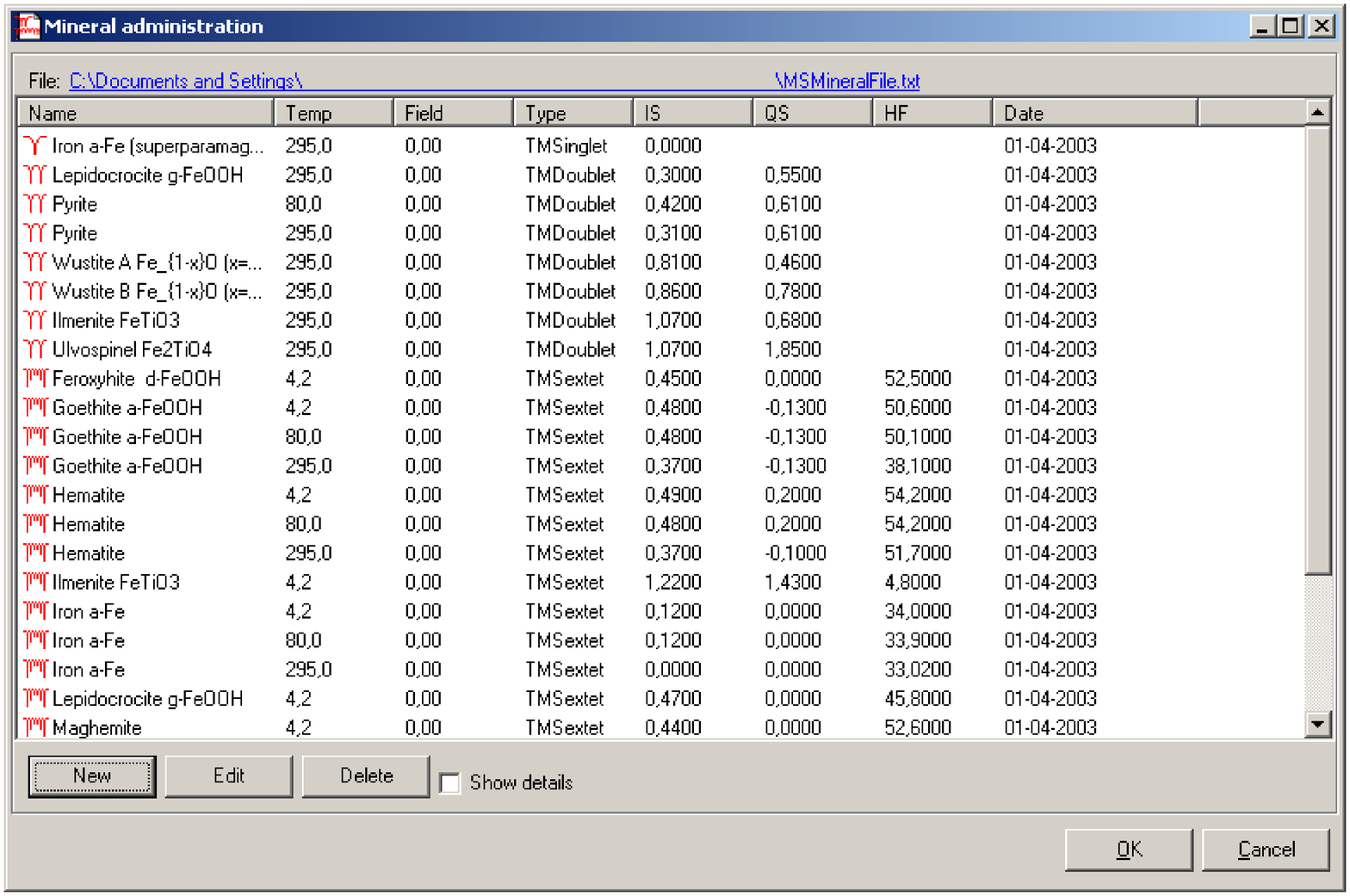}
\caption{\textit{The minerals administration window.}}
\label{fig:Minerals}
\end{figure}

The program contains a list of common and well-known minerals
(\fref{fig:Minerals}), which can be used for as starting point
fitting. The mineral data are from \cite{SM}. It is possible to edit
the listed minerals and save these modifications, and to enter new
minerals. This can be done either through the \vb{New}, \vb{Edit} or
\vb{Delete} options, or by editing the minerals file manually. The
name and location of the minerals file is displayed in top of the
window.

\section{Project planning}\label{sec:ProjectPlanning}

In the following we discuss the properties of the program as used
for analysis of a transmission/absorption spectrum. However, the
discussion is completely valid also for scattering/emission
spectroscopy, since there will be only few details differentiating
these.

\subsection{Line shapes}

Ideally each \mssb\ absorption line has a \Lorentzian{} line shape, (see \tref{tab:lineshapes}). However, there are physical effects that might
disrupt or distort the ideal case. For these cases other line shapes are necessary.

In practice most absorbtion lines are \Lorentzian{}s, which may be
sligthly smeared by a \Gaussian{} due to for instance temperature
fluctuations, leading to the Voigt line shape. The Voigt line shape
is a convolution of a \Lorentzian{} and a \Gaussian{} and can be
expressed as \cite{VoigtFunction}
\beq
L_{Voigt}(v)=\int^\infty_{-\infty}L_{Gau}(v')L_{Lor}(v-v')dv'.
\eeq

However, this can not be solved analytically, and therefore several
approximations to this profile exist. We have implemented three of
these: \PseudoVoigt{}, \PseudoLorentz{} and \PearsonVII{}, which are
listed in \tref{tab:lineshapes}. Especially the \PearsonVII{} line
shape is difficult to implement, since it involves the
implementation of the Gamma function \cite{GammaFunction}, which was
implemented using Stirlings approximation
\cite{StirlingApproximation}
\beq
\Gamma_{Stirling}(z)\cong \sqrt{\frac{2\pi}{z}} \Bigg{(} \frac{z}{e} \sqrt{z\sinh\frac{1}{z}+\frac{1}{810z^6}}  \Bigg{)}^z,
\eeq
which is accurate to a least 5 digits for z>1.

\begin{figure}[!htb]
\centering
\includegraphics[width=0.4\textwidth]{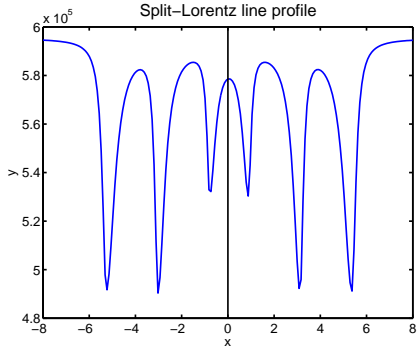}
\caption{\textit{ An example of a spectrum with \SplitLorentzian{}
line profiles. The outer slope of an absorption line is defined as
the slope farthest from the center of the component. This of course
only applies to doublets and sextets.}} \label{fig:SplitLorentz}
\end{figure}

The \SplitLorentzian{} (see \fref{fig:SplitLorentz}) line shape is
commonly used for samples in which small-particle effects or
interactions distort the line shapes in such a way that the outer
slope is steeper than the inner slope. The \SplitLorentzian{} line
shape is made up of the outer side using one \Lorentzian{} line
shape, and the inner side using another shape. The amplitude and
center of the two line shapes are kept equal, but the widths vary.
The relationship between the widths is usually controlled by
defining a width parameter and an asymmetry parameter. The asymmetry
parameter is usually known as $b$.

When using the \SplitLorentzian{} line shape in doublets and sextet (see \sref{sec:FitComponents}) the line shapes are mirrored with respect to
the center of the component, and the left side of a line shape is identical to the right side of its mirrored twin.

The line shapes implemented for the \mssb\ analysis and the properties of these are presented in \tref{tab:lineshapes}.

\begin{table}[htb]
\begin{tiny}
  \centering
\begin{tabular}{|l|c|c|c|} \hline

  Name & Formula & Area & Additional \\
    & & & parameters \\ \hline
  \vrule height 25pt width 0pt

   \Lorentzian{} & $L_{Lor}(v)=I_0\big{(}1+\big{(}\frac{2(v-v_0)}{W} \big{)}^2\big{)}^{-1}$ &  $A_{Lor}=\frac{I_0W
  \pi}{2}$&-
  \\ \hline

  \vrule height 25pt width 0pt
  \Gaussian{} & $L_{Gau}(v)=I_0 \exp \big{(}-\big{(}\frac{2(v-v_0)}{W} \big{)}^2 \big{)}$ &  $A_{Gau}=\frac{I_0W
  \sqrt{\pi}}{2}$&-
  \\ \hline

\vrule height 25pt width 0pt
  \SplitLorentzian{} &   $L_{S-L}(v)= \bigg{\{}
\begin{array}{cc}
    L_{Lor}(W_1) &, \quad v<v_0 \\
    L_{Lor}(W_2) &, \quad v\geq v_0
  \end{array}
    $
  & $A_{S-L}=\frac{1}{2}\sum_{i=1}^2A_{Lor}(W_i)$&
$\begin{array}{c}
  b\equiv \frac{W_1}{W_2}-1 \\
  W_1\geq W_2
\end{array}$ \\
  \hline
 \vrule height 10pt width 0pt
  \PseudoVoigt{} & $L_{Pseu}(v)=\eta \cdot L_{Lor}+ (1-\eta)\cdot L_{Gau}$ &  $A_{Pseu}=\eta \cdot A_{Lor}+ (1-\eta)\cdot A_{Gau}$& $\eta\in[0;1]$
  \\ \hline
 \vrule height 25pt width 0pt
  \PseudoLorentz{} & $L_{PLor}(v)=I_0\big{(}1+\big{|}\frac{2(v-v_0)}{W}\big{|}^{2+\alpha} \big{)}^{-1}$ &  $A_{PLor}=\frac{I_0W
  \pi}{2}\frac{2sin(\frac{\pi}{2+\alpha})}{2+\alpha}$& $\begin{array}{c} \alpha \in [0;1.45]\\\alpha=0 \quad Lorentz \end{array}$
  \\ \hline
 \vrule height 25pt width 0pt
  \PearsonVII{} & $L_{PVII}(v)=I_0 \big{(}1+\big{(}\frac{2(v-v_0)}{W} \big{)}^2\frac{1}{m}  \big{)}^{-m}$ &  $A_{PVII}=\frac{I_0W\sqrt{m\pi}}{2}\frac{\Gamma(m-\frac{1}{2})}{\Gamma(m}$&
  $\begin{array}{c} m\geq 1\\
  m=1\quad Lorentz\\
  m\rightarrow\infty \quad Gauss \end{array}$
  \\ \hline

\end{tabular}
\end{tiny}
  \caption{\textit{Line shapes used for \mssb\ analysis. Reproduced from \cite{Hansen}.}}\label{tab:lineshapes}
\end{table}

\subsection{Fit components}\label{sec:FitComponents}

The three fundamental \mssb{} fitting components are the singlet,
the doublet and the sextet with 1, 2 and 6 absorption lines,
respectively. All of these components are characterized by their
isomer shift (\fb{IS}).

The splitting of the doublet depend on the quadrupole interaction,
and is in the doublet quadrupole splitting (\fb{QS}).

The singlet, doublet and the sextet have furthermore the line shape
a parameter ($b$, $\eta$, $\alpha$ and $m$ in \tref{tab:lineshapes})
in common. The \SplitLorentzian{}, \PseudoVoigt{}, \PseudoLorentz{}
and \PearsonVII{} line shapes use this property.

The magnetic Zeeman interaction, the magnetic hyperfine field
(\fb{HF}), is only used in the sextet, and needs only to be
implemented in the sextet. For sextets this software - presently -
can handle only spectra in which the magnetic Zeeman interaction is
dominant, so that the quadrupole interaction can be treated as a
small perturbation on the magnetic interaction. In the case of the
sextet \fb{QS} indicates the quadrupole shift in contrast to the
doublet quadrupole split.

However, since the properties \fb{HF}, and \fb{QS}, are the only
properties which do not apply to all components, we have provided an
interface to them in all components, but disabled them where they
are not needed. As mentioned, we do not use the complete
spin-Hamiltonian for calculations of the line positions in the, but
rather approximate expressions, which however, have been quite
sufficient in most all cases.

The singlet consists of a single line, and the position of the
single line is the same as the isomer shift. The expression for the
singlet is given by

\beq
y_i = L(v_i-v_{0}),
\eeq
where $L$ is the line shape function providing the value of $y_i$ as
a function of $v_i$, and $v_{0}$ is the value of the isomer shift.

The doublet consists of two lines, and the centers of the lines are placed at a distance of \fb{QS} apart, around the isomer shift. The
expression for the doublet is given by
\beq
y_i = L(v_i-v_{0}-\frac{v_{0}}{2})+L(v_i-v_{0}+\frac{v_{QS}}{2}),
\eeq
where $v_{QS}$ is the value of the quadrupole splitting.

The sextet consists of six lines, and the centers of the lines are
placed at positions dictated by the isomer shift, quadrupole shift
and hyperfine field.

The expression for the sextet is given by
\beq y_i = \sum_{i=1}^6
L(v_i-v_{0}+(-1)^{(1+\delta_{2345})}\frac{v_{QS}}{2}+(-1)^{(1+\delta_{456})}k_{q,w}\cdot
B_{HF}),
\eeq
where $\delta_{mnkl}$ is a delta function, which is 1 when $i$ has
one of its values. $k_{q,w}$ is a \mssb\ proportionality factor,
composed of several physical constants. The three values that apply
are $k_{1,6}$, $k_{2,5}$, $k_{3,4}$ for the paired lines. $B_{HF}$
is the hyperfine field. The $k_{i,j}$ values are calculated from the
general expression for the sextet energy levels, which is given by
\beq
\label{eq:mLevelEnergy} E_m=-g \beta_n m B,
\eeq
where $g$ is the \lande\ factor, $\beta_n$ is the nuclear magneton,
and $m$ is the m-quantum number. The energy splitting is calculated
by
\beq
|\Delta E_{(i,j)}| = |\Delta E_i|+|\Delta E_j|=2|\Delta E_i|,
\eeq
where $(i,j)=(1,6), (2,5), (3,4)$. Applied to the line pairs (1,6),
(2,5), (3,4) then energy splitting is calculated from
\begin{eqnarray}\label{eq:g_factor_label} |\Delta E_{(i,j)}|
&=&-2(\beta_nB(g_e\frac{3}{2}-g_g\frac{1}{2})) = \beta_n B \cdot
g_{ij},
\end{eqnarray}
where $g_g=0.181208$ and $g_e=-0.10355$ are the g-factors for the
ground state and the excited state respectively, values from
\cite{SM}, and the $g_{ij}$ factors,
    \begin{eqnarray}
    g_{16}&=&-2(g_e\frac{3}{2}-g_g\frac{1}{2}) = 0.491858 , \\
    g_{25}&=&-2(g_e\frac{1}{2}-g_g\frac{1}{2}) = 0.284758 , \\
    g_{34}&=&-2(g_e\frac{-1}{2}-g_g\frac{1}{2}) = 0.077658 , \\
    \end{eqnarray}
are g-factors extracted from the g-factor for the excited and ground
state and $E_0$ is the transition energy from $I=\frac{3}{2}$ to
$I=\frac{1}{2}$ for \Fe\ \mssb\ spectroscopy.

The actual calculation being performed is
\beq
v_{HF}=g_{ij}\beta_n B_{HF} \cdot \frac{c}{2 E_{0}},
\eeq
using unit conversion this becomes
\begin{eqnarray}
v_{HF}&=&g_{ij} B_{HF} \frac{\beta_n\cdot c \cdot k}{2E_{0}e}\\
 &=&g_{ij} B_{HF}\frac{5.0505\ee{-27}\,{\rm J/T} \cdot 3.0\,\ee{11}\,{\rm mm/s}}{2\cdot 14.41\,{\rm keV}\cdot 1.6022\ee{-19}{\rm J/eV}} \\
&=&g_{ij} B_{HF}\cdot 0.32794\,{\rm mm/sT},
\end{eqnarray}
where $e$ is the elementary charge. Now the $k_{q,w}$'s can be
calculated as
\begin{eqnarray}\label{eq:mssbConstants}
k_{1,6}&=&1.61299\ee{-1}\,{\rm mm/sT},\\
k_{2,5}&=&9.33835\ee{-2}\,{\rm mm/sT},\\
k_{3,4}&=&2.54672\ee{-2}\,{\rm mm/sT}.
\end{eqnarray}

\subsection{Fitting spectra}
The program will analyze \mssb\ spectra by fitting a model set by
the user, and it will report the result of the fit.

Before any analysis can commence, a background level has to be
established/calculated. The background level is calculated as the
mean of the value of the outermost 8 channels on each side of the
spectrum. Transmission spectra contain absorption lines, which have
fewer counts than the background level. The expression for
components provided above has to be subtracted from the background
to produce a model data series.

The program can also be used for analysis of reflection spectra.
Reflection spectra also contain a background level, to which the
reflection lines are added. To produce a fitting model, the fitting
components are added to the background level.

\subsection{Calibration spectra}\label{sec:CalibrationSection}

The program derives the calibration parameters of the spectrometer
from automatic analysis of calibration spectra. The found
calibration parameters are then used in the analysis of spectra
obtained under circumstances identical to those of the calibration
spectrum. In this program version (\ProgramVersion) only calibration
of linear velocity profiles are implemented. We are well aware that
sinusoidal velocity profiles are common in the \mssb\ community, and
a later version may be adapted for spectra obtained in the mode.

Contrary to a normal spectrum for \mssb\ analysis, a calibration spectrum is analyzed unfolded with the channel numbers used as reference.

The calibration spectrum usually consists of the spectrum of a thin
iron ($\alpha$-Fe) foil (12.5\,$\mu$m) at 295\,K. But other
calibration materials can also be used, such as stainless steel and
other well characterized iron compounds.

The unfolded calibration spectrum consists of a background level,
and 12 absorption lines on the background level
\fref{fig:CalBeforeInsComp}. The calibration is used for finding 3
unknown parameters of the experimental setup, and to provide
information on any imperfections of the source/drive system i.e.
increased line widths, nonlinearity etc. The three unknown
parameters are listed below:
\begin{description}
    \item[The folding channel] is the channel (integer) around which the
    data channels are folded. Operationally (for historic reasons and backwards compatibility) it is defined as the channel number, which is
    added to channel number 1, e.g. it is the double of the actual folding channel.
     In a 512 channel setup, it has typically a value of 510-514. It is found by
    calculating the mean of positions of the lines (1,12), (2,11),
    etc.
    \item[The zero velocity channel] is defined as the detector channel position
    (decimal, 5 digits), of the symmetry center of the 295\,K spectrum of a thin foil of $\alpha$-Fe.
    It is found in each of the spectrum halves separately. It is found by locating the mean center of the lines
    (1,6), (2,5) and (3,4) pairwise, and correspondingly for the
    second half of the spectrum, and then finding the common
    center.

    \item[The calibration constant] is the link between
    the channel number and the source velocity. It is found as the
    mean value of the known hyperfine field of the absorber
    material ($B_{HF}=33.02$\,T for $\alpha$-Fe at 295\,K) divided by the
    distances between the transition pairs, $(i,j)={(1,6),(2,5),(3,4)}$, and scaled by the \mssb\
    constants \eref{eq:mssbConstants}.
    The exact formula is
    \beq
       c_{cal}=\frac{1}{6}\sum \frac{g_{ij} \beta_n B_{HF}
       }{E_0\cdot c}\bigg{(}\frac{1}{P_j-P_i}+\frac{1}{P_{j+6}-P_{i+6}}\bigg{)},
    \eeq
    where $P_i$ is the channel position of the \textit{i}'th absorption
    line.
\end{description}

The background line has a shape that is dependent on the mode of
operation of the spectrometer (e.g. frequency and velocity range)
and geometry of the experimental setup. The radioactive source is
placed inside a collimator, and since the source is oscillating with
near constant acceleration, the solid angle seen by the source will
vary as a function of position. The radiation detected is
proportional to the angular area. The baseline shape is approx. that
of two parabolas in succession with positive and negative $a$
values:
\begin{eqnarray}
A &=&+a\cdot(v-c_M)^2-b(v-c_M)+c, \quad 1\leq v \leq
c_M, \\
A &=&-a\cdot(v-c_M)^2-b(v-c_M)+c, \quad c_M \leq v \leq N,
\end{eqnarray}
where $A$ is the angular area, $c_M$ is the mirror channel, $v$ is the channel number, $N$ is the number of channels, and $a$, $b$ and $c$ are
the parabola parameters.

Besides the above calculations, the calibration is able to find the
12 absorption lines ($^{57}$Fe), and fit these and the baseline to
find the calibration constants.

As the equipments available during the development only used a
linear velocity drives, only this case has been implemented in the
program, but this is easily expanded, when the apparatus data are
available.

\section{The program structure}\label{sec:ProgramStructure}
The program is built entirely on object oriented technology (OOT).
There are several advantages in using OOT:
\begin{itemize}
  \item Inheritance makes it easy to create a child class, which inherits
        most of its properties and methods from its ancestor, but
        introduces some new, or changes the internal behavior in some
        areas.
  \item Isolation makes it easy to correct errors or undesirable
        behavior, without affecting other parts of the program, thereby
        minimizing program errors. Furthermore it makes it easy to
        extend functionality.
\end{itemize}

There are also disadvantages in choosing OOT:

\begin{itemize}
  \item Processing speed will in most cases be lower than that of
a procedural program since the overhead will be larger, and the amounts of data manipulated in most cases will be larger.
  \item The implementation process will take longer and the source code
will be larger, since similar behavior may have to be implemented several times in order to avoid code shared between classes.
\end{itemize}

The program was implemented using Borland Delphi 5, 6 and 7, and
should be executed on a PC running Microsoft Windows 2000 or
Microsoft Windows XP. During implementation the built-in classes of
Delphi have been used as programming base. For increased flexibility
the program is designed to be a multiple document interface (MDI)
program, meaning that it would be possible to work with several
'documents' (spectra) simultaneously.

\subsection{The fitting form} 
\begin{figure}[!htb] \centering
\includegraphics[width=0.6\textwidth]{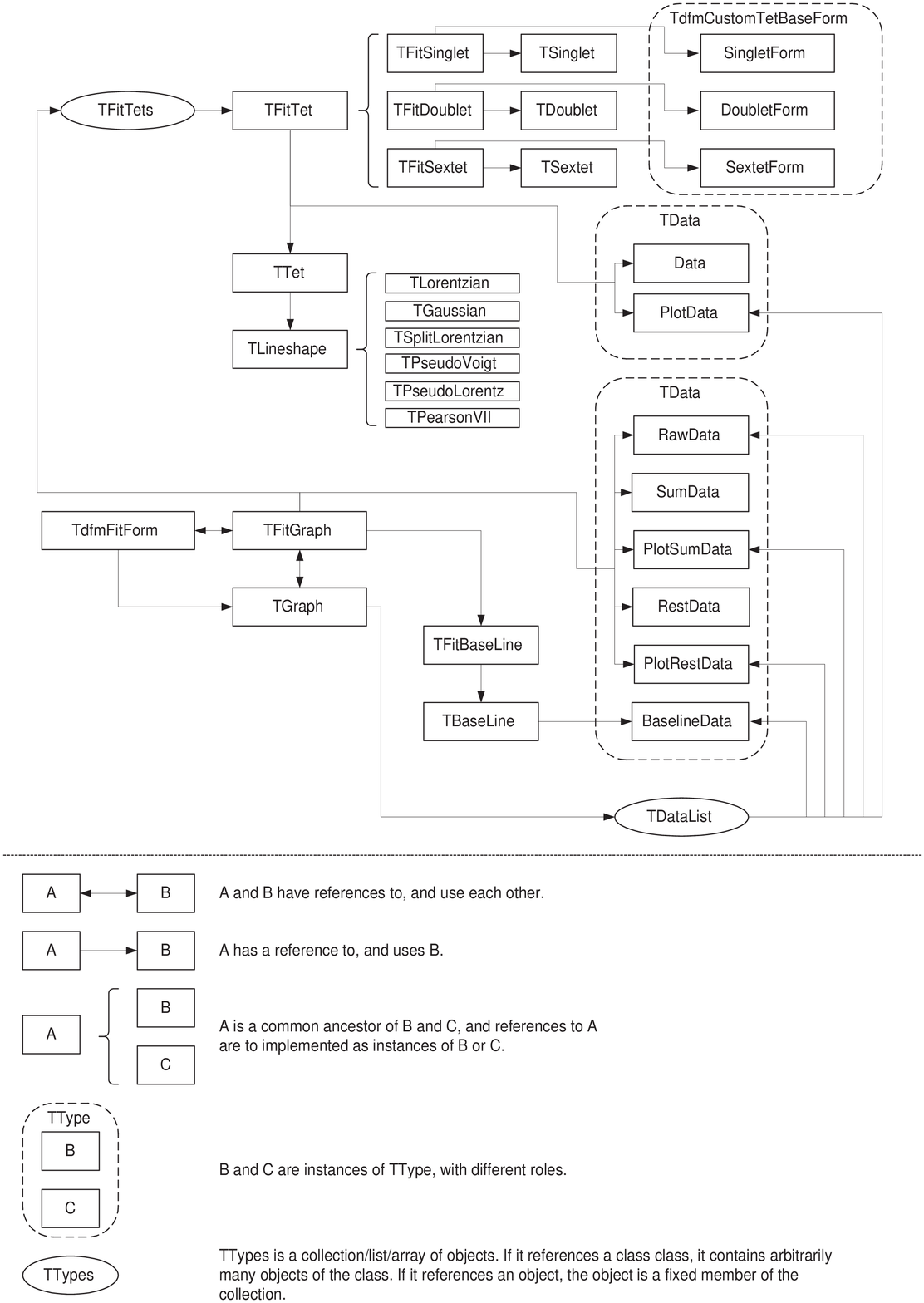}
\caption{\textit{The logical/workflow diagram of the \pb{TdfmFitForm} class, used for fitting \mssb\ spectra. The upper part is the
logical/workflow diagram, and the symbol explanations are shown in the lower part. }} \label{fig:FitForm}
\end{figure}

The \mssb\ fitting form is
implemented as the class \pb{TdfmFitForm} (see \fref{fig:FitForm}).
It is the graphic user interface, through which the user makes input
to and receives output from the spectrum analysis. It contains two
very important classes \pb{TGraph} and \pb{TFitGraph}.

After the FitForm has been created and initialized, it loads a
data-file (which the user has chosen), and calculates the \mssb\
spectrum background level. The spectrum is folded during loading.

Whenever a change has been made to one of the fitting components, or
a new fitting component has been inserted, several calculations have
to be performed:
\begin{itemize}
  \item The sum data, \pb{SumData}, which contains
the sum of the fitting components is calculated as
  \beq
  S_i = \sum_{j=1}^{N_{tet's}} T_{j,i},
  \eeq
where $S_i$ is the sum at the i'th point, $T_{j,i}$ is the value of
the j'th fit component (singlet, doublet, sextet), at the i'th
point.
  \item The rest
data, \pb{RestData}, which contains the difference between the
spectrum data and the sum is calculated as
\begin{eqnarray}
R_i &=& D_i - (B_i-S_i) = D_i-M_i \quad {\rm for\ transmission\ spectra,}\\
R_i &=& D_i - (B_i+S_i) = D_i-M_i \quad {\rm for\ scattering\
spectra,}
\end{eqnarray}
where $R_i$ is the rest at the i'th point, $D_i$ is the i'th data
point of the spectrum data, $B_i$ is the i'th point in the baseline
and $M_i$ is the model data at the i'th point.

  \item The $\chi^2$, \pb{ChiSqr}, and $\tilde{\chi^2}$ which
are indicators for difference between the model and the data, and
used for fitting, is calculated. The misfit is also given as an
output parameter. It can be summarized mathematically to
\beq
  \chi^2 =  \frac{1}{n_{valid}} \sum_{i=1}^{n_{valid}} \frac{R_i^2}{D_i}=\frac{1}{n_{valid}} \sum_{i=1}^{n_{valid}}\frac{(D_i-M_i)^2}{E_i^2},
\eeq
since $E_i=\sqrt{D_i}$ and $n_{valid}$ is the number of valid
channels in the spectrum. Furthermore we calculate the reduced
chi-squared $\tilde{\chi}^2$ by
\beq
  \tilde{\chi}^2 =  \frac{1}{n_{valid}-n_{free}} \sum_{i=1}^{n_{valid}} \frac{R_i^2}{D_i},
\eeq
where $n_{free}$ is the number of free fit parameters. As a goodness
of fit parameter we use the misfit $m$ \cite{Waychunas} defined by
\beq
m=\frac{n(\chi^2-1)}{\sum_{i=1}^{n_{valid}}\frac{(B_i-D_i)^2}{D_i}}.
\eeq

\end{itemize}

Two fitting algorithms, random walk \cite{Hansen} and amoebe
\cite{NumRecipes}, have been implemented in the fitting form. First
we will go through the random walk algorithm.

\begin{figure}[!htb]
\centering
\includegraphics[width=0.5\textwidth]{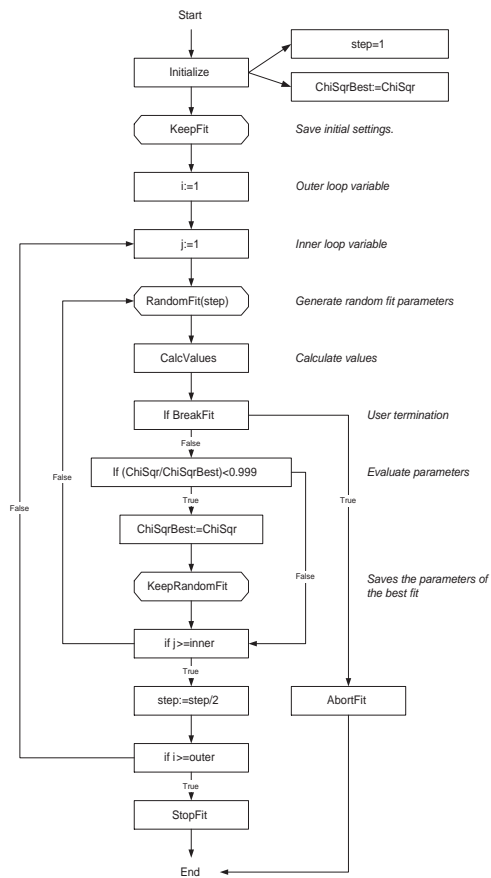}
\caption{\textit{A graphical illustration of the RandomFit fitting algorithm. The steps marked with octagonal boxes indicate that this step
iterates through all TFitTet's in the fit.}} \label{fig:RandomFit}
\end{figure}

The random walk algorithm is a variant of the "simulated
annealing"-algorithm used in many areas of physics. It differs
however, in that the tunneling possibility has been eliminated. The
\pb{RandomFit} algorithm is presented graphically in
\fref{fig:RandomFit}. The algorithm is based on two loops, the outer
and the inner. The core of the algorithm is thus run
  \beq
  N=n_{outer}\cdot n_{inner}
  \eeq
times. The algorithm starts with setting the variable \pb{step} to
1. Each time the outer loop is traversed once, \pb{step} is set to
one half of its previous value. The philosophy is to find the
hitherto best fit for each time the outer loop has been traversed,
and thereafter to narrow the interval in which the parameters are
allowed to vary, before traversing to the outer loop again.

The algorithm is based on the assumption that the best fit after
$n_{inner}$ guesses, lies in the vicinity of the best fit. The
fitting parameters all have an interval to which they are restricted
during fitting. Using the result of the sum
\beq \sum_{i=0}^{\infty} 2^{-i}=2, \eeq this can be
maintained/fulfilled. If $\Delta$ is the fitting variation interval of a fitting parameter, the following formula \beq \sum_{i=0}^{n_{outer}-1}
\Delta\cdot 2^{-i}<2\cdot \Delta \eeq provides a way to implement our version of the random walk algorithm without violating the restrictions:
\beq \sum_{i=0}^{n_{outer}-1} 1/2\cdot \Delta\cdot 2^{-i}<1/2\cdot2\cdot \Delta=\Delta. \eeq Before the first traversation of the inner loop,
the variation intervals are therefore set to half of the original, corresponding to the formula above. This guarantees that the fit parameters
are kept within the allowed fit interval.

The amoebe algorithm \cite{NumRecipes} works in many areas much like
the random-walk algorithm. It has an outer loop, which repeats
$n_{outer}$ times, or until the improvements are no longer
significant. The inner loop basically works same way as in random
walk.

\begin{figure}[!htb]
\centering
\includegraphics[width=0.5\textwidth]{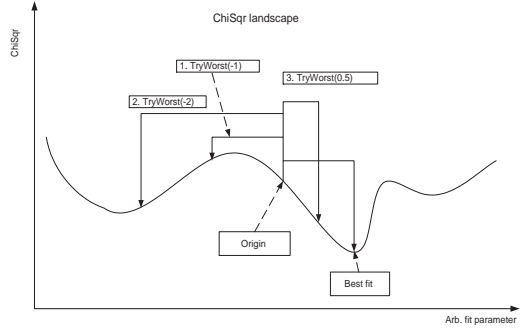}
\caption{\textit{A graphical illustration of the preprogrammed steps in the amoebe fitting algorithm.}} \label{fig:AmoebeTrix}
\end{figure}

Besides the evaluation of the improvements mentioned, the difference
between the random walk and amoebe is that after the best fit has
been found using the inner loop, some preprogrammed variations of
the best fit are tried. For all paramters the following is tried
individually: Try to the fit parameter value in the opposite
direction, with the same amount. If this gives a better fit, it
tries to shift the parameter value a step further in the same
direction, to see if it should give an even better fit. This tests
if the point of origin was a local maximum, and if the fit has found
the worst local minimum. If so, the algorithms tries if a further
step in the same direction gives a better fit.

If the opposite step did not yield a better result, the algorithm tests to see if it has stepped too far to find the currently best fit, and
tries to see if there is a better point halfway between the origin and the currently best fit. If all of these attempt do not result in a better
fit the \pb{step} variable is divided by 2, and the content of the outer loop is traversed again. If a better fit is found \pb{step} is not
changed. The preprogrammed steps are illustrated in \fref{fig:AmoebeTrix}.

The outer loop will terminate when it has been traversed 20 times,
terminated by the user or when the \pb{rtol} variable is larger than
the \pb{ftol}. \pb{ftol} is a measure of the minimum relative
improvement required, for improvements to be considered significant.
\pb{rtol} is a measure of the improvement of the best fit compared
to the worst fit in the current traversation of outer loop.
\pb{rtol} is calculated from:
 \beq
 r_{tol}=2\frac{|\chi^2_{worst}-\chi^2_{best}|}{|\chi^2_{worst}|+|\chi^2_{best}|}.
 \eeq

\begin{figure}[!htb]
\centering
\includegraphics[width=0.5\textwidth]{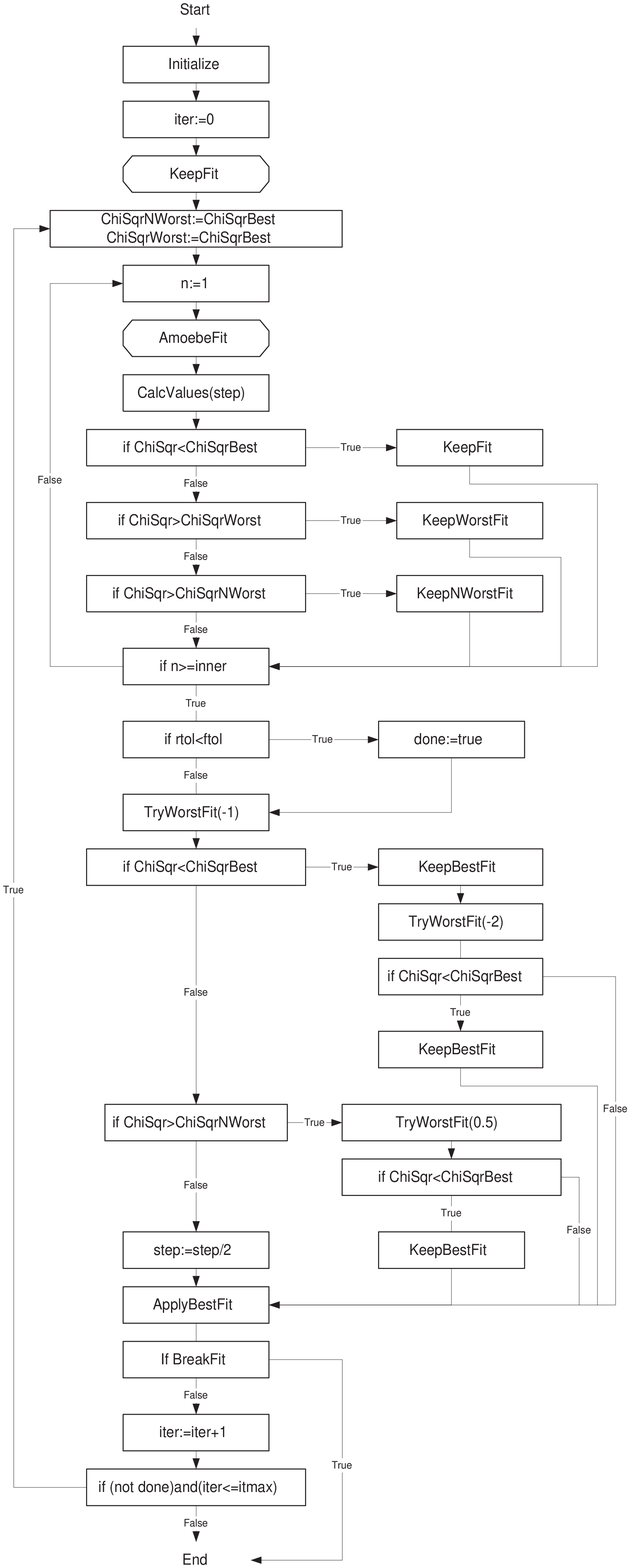}
\caption{\textit{A graphical illustration of the \pb{AmoebeFit} fitting algorithm. The steps marked with octagonal boxes indicate that this step
iterates through all \pb{TFitTet}'s in the fit.}} \label{fig:AmoebeFit}
\end{figure}

The source code is presented graphically in \fref{fig:AmoebeFit}.

%-------------------------------------------------------------------------------The calibration form

\subsection{The calibration form}

\begin{figure}[!htb]
\centering
\includegraphics[width=0.6\textwidth]{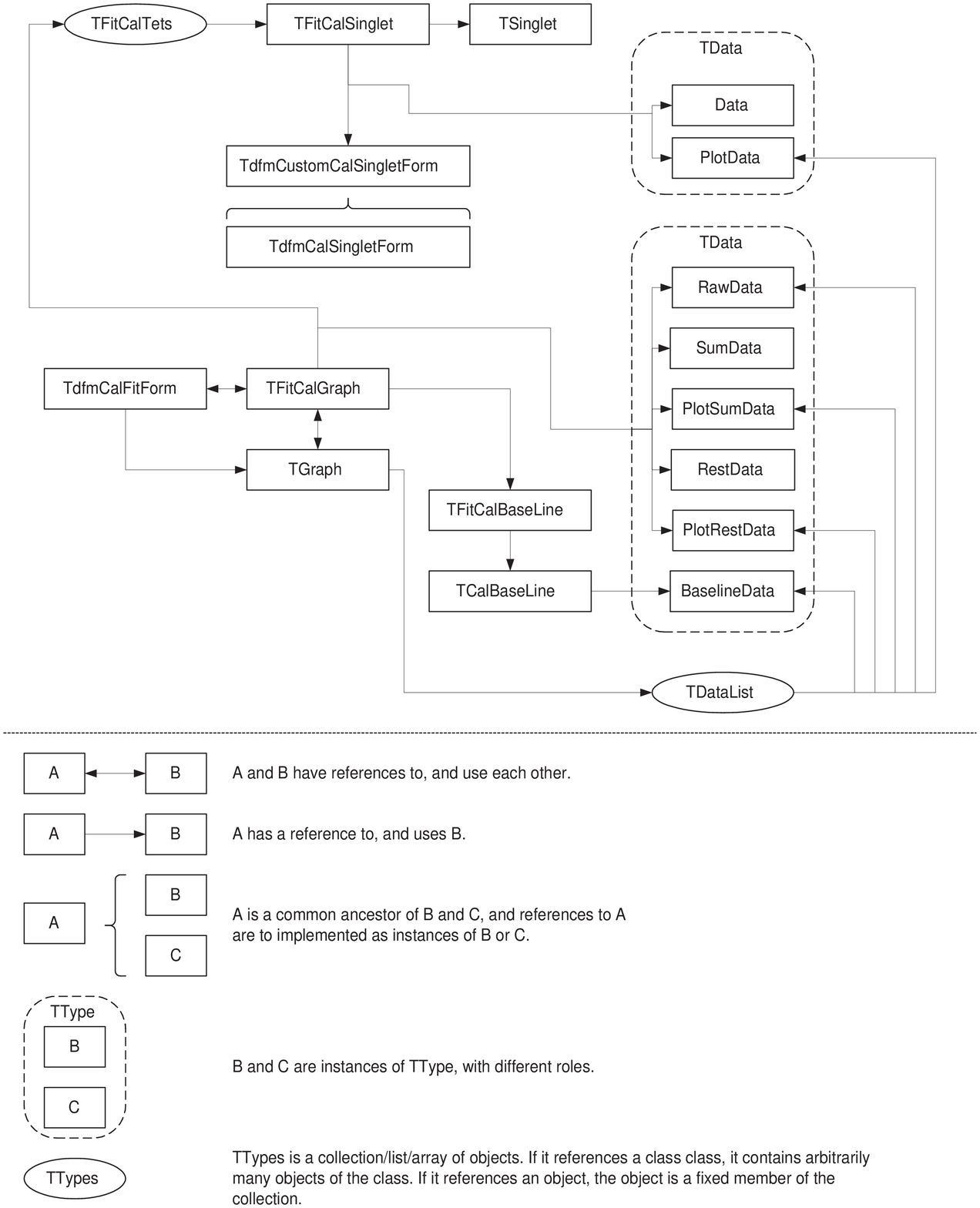}
\caption{\textit{The logical/workflow diagram of the \pb{TdfmCalForm} class, used for fitting \mssb\ spectra. The upper part is the
logical/workflow diagram, and the symbol explanations are shown in the lower part. }} \label{fig:CalForm}
\end{figure}

The \mssb\ calibration form is the graphic user interface, through
which the user makes input to and receives output from the
calibration process. Generally it resembles the fitting form in many
points, and the description of common features will not be repeated.
The data files are, however, loaded unfolded.

Whenever a change has been made to one of the fitting parameters,
the same calculations as in the fitting form have to be performed.
\pb{SumData}, \pb{PlotSumData}, \pb{RestData} and \pb{PlotRestData},
are calculated in the same way. The only difference with regards to
the fitting form is the variable baseline in the calibrations form,
which has to be taken into account.

Before the calibration process can commence, the fitting components
have to be inserted, and the baseline parameters defined. The
baseline is initially fixed, but is allowed to vary during the
calibration. The 12 components are inserted at the 12 local extrema
with the largest 'magnitude'.

The local minima are found by traversing through the data points of
the data, and for each data point it checks whether the current
point is a local extremum by comparing it to its neighboring points.
If the point is a local extremum the magnitude is calculated as
  \beq
  M_{pivot}=\frac{y_{i+2}+y_{i+1}+y_{i-1}+y_{i-2}}{4}-y_i,
  \eeq
where $y_i$ is the i'th point. After finding the all local extrema,
the list of extrema is sorted, so the local extrema with the largest
magnitude are placed first in the list, and the predefined number of
largest local extrema are returned, and used for inserting the
fitting components.

The random walk in the calibration form is very similar to that of the fitting form. The only difference lies in that the inner loop fits the
components parameters when the inner loop counter \pb{j} is even and the baseline parameters when it is odd. This is done since experiments
during programming shows that this improves the intelligence and speed of the calibration procedure.

\section{Conclusion}
We have presented the functionality and the basic principles behind
the \fit\ program. We believe that by the program we have made
available a valuable tool for general simple \mssb{} analysis of
complex samples.

\section{Acknowledgements}
We would like to thank Preben Bertelsen and Kristoffer Leer of the
Mars/\mssb{} group at The Niels Bohr Institute at Copenhagen
University for testing and suggestions. From Ris\o{} National
Laboratory, Roskilde, Denmark we would like to thank Peter Kj\ae{}r
Willendrup, Kim Lefmann for comments, and Luise Theil Kuhn for
comments, suggestions and testing.

\section{Additional information}\label{sec:AdditionalInformation}
The program is available from the \fit\ homepage at
\cite{FitWebpage}. There is a forum section, a FAQ, as well as a
short manual available.

\clearpage
\appendix
\section{Examples}\label{sec:Expamples}
There are a few .exp-files included in the installation package.
These are test spectra, which can be used for practicing. A
description of the contents of the files is given below:

\begin{itemize}
 \item \vb{A10000.exp}, \vb{A10000el.exp},  \vb{A10000er.exp} are
 simulated/generated testing spectra. They contain a perfect
 $\alpha$-Fe sextet on both, right and left side respectively.
 \item  \vb{A10001.exp} is a spectrum of soil from Salten Skov,
 \AA{}rhus, Denmark.
 \item  \vb{A10002.exp} is an $\alpha$-Fe transmission calibration
 spectrum from a Fe foil.
 \item  \vb{B10001.exp} is an $\alpha$-Fe scattering calibration
 spectrum from a Fe foil.
 \item  \vb{B10105A.exp} \vb{B10001.exp} is an $\alpha$-Fe scattering calibration
 spectrum from a 25\,$\mu$ Fe foil, with an perpendicularly applied magnetic
 field.
 \item  \vb{B10106a.exp} is an $\alpha$-Fe scattering calibration
 spectrum from a 25\,$\mu$ Fe foil.
 \item \vb{HA668.exp}, \vb{HA679.exp}, \vb{HA869.exp}, \vb{HA900.exp}, \vb{HB300.exp}, \vb{HB506.exp},
 \vb{HB522.exp} and \vb{HE0195.exp} are real spectra of various
 samples, both natural and synthesized.
\end{itemize}

\bibliographystyle{elsart-num}
\bibliography{fitarticlereferences}
\end{document}